# Hybrid Multi-Walled Carbon Nanotube TiO$_2$ Electrode Material for Next Generation Energy Storage Devices

Sydney Marler

*Sherman E. Burroughs High School,*

*500 E. French Avenue, Ridgecrest, CA 93555*

Mentor: Dr. Jason Li, Asylum Research, Santa Barbara, CA

*Abstract* - Current supercapacitors present several distinct limitations that severely inhibit the efficiency, power, and electrical capacitance of energy storage devices. Supercapacitors present an exciting prospect that has countless applications in renewable energy storage and modern day electronic devices. In recent years the exciting development of carbon nanotubes (CNTs) has presented an advantage in electrode development. CNTs, however beneficial for their increased electrode surface area, have severe limitations regarding conductivity and electrode density. Creating a nanocomposite hybrid out of a transition metal-oxide and carbon nanotube array would help the current limitations of the modern supercapacitor. TiO$_2$ was chosen for its common occurrence in everyday materials and promising capacitance levels. A multi-walled carbon nanotube array was grown on a SiO$_2$ precursor via CCVD. The transition metal oxide was then deposited via RF Sputtering methods to a MWCNT array. Recharge tests and characterization were conducted using scanning capacitance microscopy (SCM). While these tests are preliminary, this novel hybrid electrode represents an exciting prospect for the future of the efficiency of electrochemical energy storage as well as an advance towards a future of providing inexpensive energy storage solutions around the world.

Introduction

Providing efficient and powerful energy storage is a key step to revolutionizing many modern technologies, particularly in the field of renewable energy. Recently, the application and integration of nanotechnology in electrochemical capacitor electrodes has attracted a considerable amount of attention in the scientific community (Baughman et al 2002). Preliminary tests using carbon nanotubes (CNTs) have shown much promise in the improvement of energy density and capacitance of ultra-capacitor electrodes (Puthukodan 2014). CNTs are cylindrical allotropes of carbon that have been applied to a wide variety of structural and nanotechnological engineering and design problems, including supercapacitors (Pan et al. 2010). However, there are several important inhibitors to the further application of electrochemical capacitance devices (Kaempgen, et al 2009; An et al 2009). The first is the limited electrode surface area; the ability of electrochemical capacitors to include higher energy storage capacity and conductivity lies in the porous surface area of the electrode (Shan et al 2007; Frackowiak et al 2000). Thus, CNTs are ideal in enhancing the total surface area. However, another problem with the energy density arises with the application of CNTs in supercapacitors. Density must then be increased in the electrode using additional materials that are conducive to conductivity without inhibiting the advantages gained by the increased surface area (Frackowiak et al 2002).

More factors in the design process of the electrochemical capacitor include the rate at which electroactive components can approach and leave the electrode surface area, the rates of side reactions, and the rate of irreversible entropic processes. The ideal capacitor device would be evaluated under the following criteria: energy density, safety, and capacitance. An ideal substrate for deposition is established as a transition metal oxide in previous literature.

The criteria used in the selection of a transition metal oxide substrate included a literature search involving the overall aspects of an ideal capacitor device. Energy density was evaluated using mathematical models; the electric current density can be calculated using Planck's constant and the charge of the electron. The total electric current in the electrode can be integrated, where G is the electric current density.

$$G_o = \frac{2e^2}{h}$$

$$I = \int G \, dS$$

*The hypothesis is that if an electrode composed of arrayed multi-walled carbon nanotubes (MWCNTs) is deposited with a nanocomposite titanium dioxide layer, then the electrode will allow for increased conductivity and energy efficiency in applications to the standard electrochemical capacitor (Xiao et al 2003). Theoretically, this will allow the electrochemical capacitor to overcome key limitations to practical application to the storage of renewable energy (May et al 1999).*

Materials

**MWCNTs-** Densely layered CNT layers that are grown via heat deposition onto a quartz substrate layer. The quartz substrate layer used has dimensions of 1" length by 1" width. The CNTs are black in color, odorless, and in a powdered form. Melting point is defined at 3,652 degrees Celsius. The MWCNTs are insoluble in water.

**$TiO_2$ Sample-** A naturally occurring transition metal oxide that is commonly found and used in manmade products.

**Quartz Substrate-** The precursor and growth substrate for MWCNTs. Dimensions of 1" length by 1" width.

**Intlvac Nanochrome Metallica Sputtering-** A non-conformal deposition process that integrates Radio Frequency Sputtering (RF Sputtering) with transition metal-oxides for electrode synthesis. Argon gas pressure is used to control sputter deposition rates within a heat control chamber. Particle energy decreases with increased argon pressure. The substrate and the chamber form an electrode, with the target substrate as the cathode. This process allows for the deposition of the transition metal oxide.

**Scanning Capacitance Microscopy (SCM)-** A characterization technique that measures the properties of thin films over the span of several microns and analyzes nanometer-by-nanometer variations in the capacitance of surface atomic structures.

Materials and Methods

The methodology of this research was divided into four phases: Design, Synthesis, Characterization, and Testing.

**Design**

A decision matrix was constructed for the CNT density, the transition metal-oxide, and the CNT Growth Substrate. Major factors included the predicted current density, electric conductivity, capacitance, and price. Multi-walled Carbon Nanotubes (MWCNTs) at a density of approximately 2.1 g/cm$^3$ while stored at 20 degrees Celsius were selected for synthesis. Titanium was chosen as the transition metal oxide substrate due to its promising conductivity, energy density, and common usage in everyday products.

A layer of multi-walled carbon nanotubes were remotely grown on a quartz substrate via combustion chemical vapor deposition (CCVD). This process involves the usage of a 1" length by 1" width SiO₂ precursor which is chemically heated, turned highly reactive, and then attached to the MWCNT target.

Two samples were chosen for characterization in order to test different deposition thicknesses and by extension, two different energy densities. 1000 Å and 50 Å thicknesses were chosen to deposit in layers of titanium dioxide onto the MWCNTs.

**Synthesis**

The MWCNT/SiO₂ substrate was synthesized using CCVD and the TiO₂ transition metal oxide precursor was prepared for deposition. The Intlvac Nanochrome Sputtering was prepared for an 18 minute sputtering cycle (Figure1 and Figure 2).

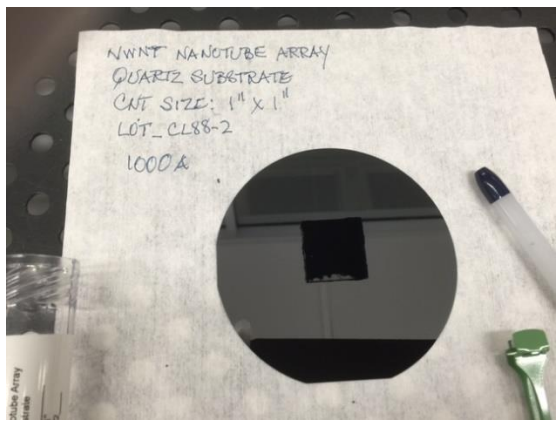

Figure 1. Sample 1 with a deposited thickness of TiO₂ onto a MWCNT/SiO₂ target. The TiO₂ is deposited via RF Sputtering using the Intlvac Nanochrome Sputtering Equipment at a thickness of 1000 Å

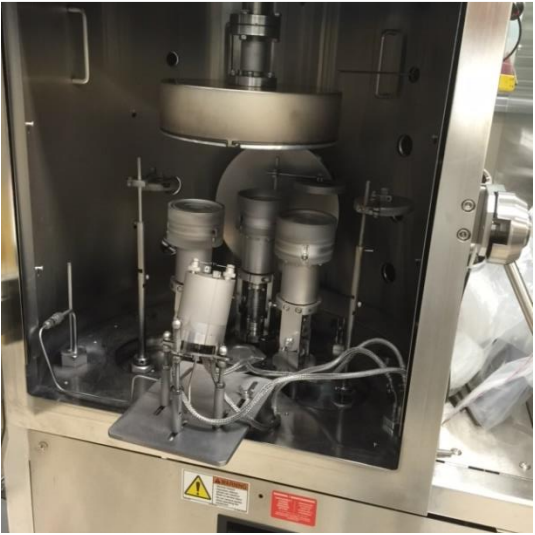

Figure 2. The Intlvac Nanochrome Sputtering device placed in "spin mode" during the cycle of the

1000 Å deposition. Each of the deposition cycles were conducted for 18 minutes.

Figure 3. The settings of the Intlvac Nanochrome Sputtering Process. The RF power, deposition time, and chamber air were variables used in deposition. The deposition time was constant for both samples,

while the RF power and argon gas pressure varied for each sample.

Both samples (1000 Å and 50 Å) were placed in the Intlvac Nanochrome Sputtering for a deposition time of 18 minutes (Figure 3).

**Characterization**

Characterization was accomplished using Scanning Capacitance Microscopy equipment (Figure 4). This technique analyzed the conductivity and energy density at the thin-layer molecular level. Tests involved the removal of small sections of each of the samples for analysis.

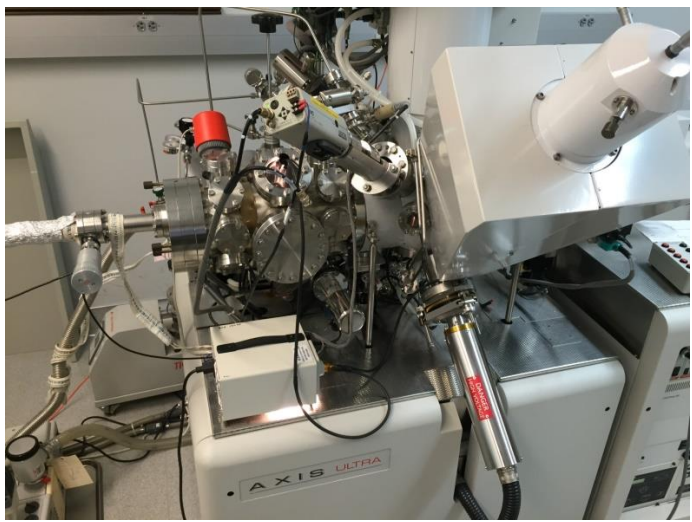

Figure 4. The AXIS Scanning Capacitance Microscopy equipment setup. Sensor Signals and output signals were set to a configuration that allowed the probe to take data from each sample.

The SCM Probe collected data on capacitance and energy density via the entrance of a data probe into the deposited sample. The sample dependent variables in this process included the sensor signals ($\Delta C/\Delta V$) and output signals ($\Delta V$).

Results

The deposition process was completed with several errors made in the 1000 Å sample due to overestimation of thickness. Several instances of peeling occurred in the samples. The sample thickness of $TiO_2$ was then changed to 742 Å, changing some of the overall calculations of predicted energy density and capacitance. This new thickness was still within a deviation of the 50 Å sample and was

considered to be still functional as an electrochemical capacitor. The 50 Å sample was deposited at 48 Å

within a reasonable machine error range.

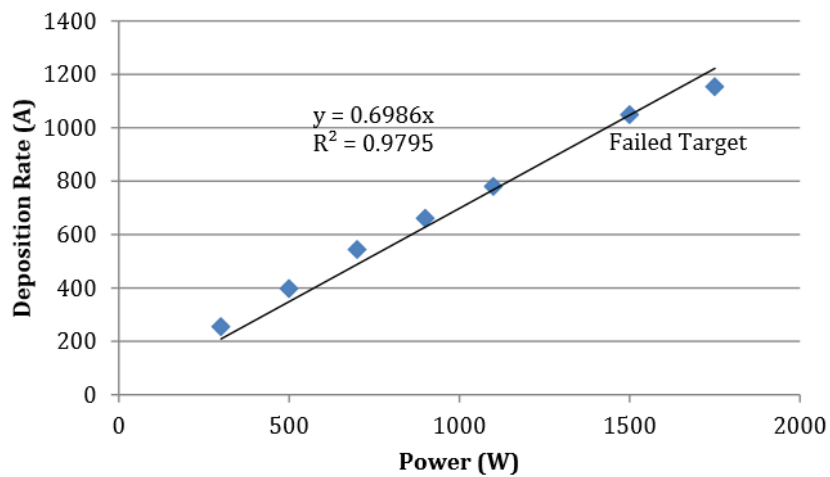

Figure 5. The deposition process recorded in power (W) as a function of deposition rate (Å). The data

proved to display linearity with additional experimentation.

Additional tests were conducted with the Intlvac Nanochrome to ensure that the deposition results followed along with accepted values reported in literature (Figure 5).

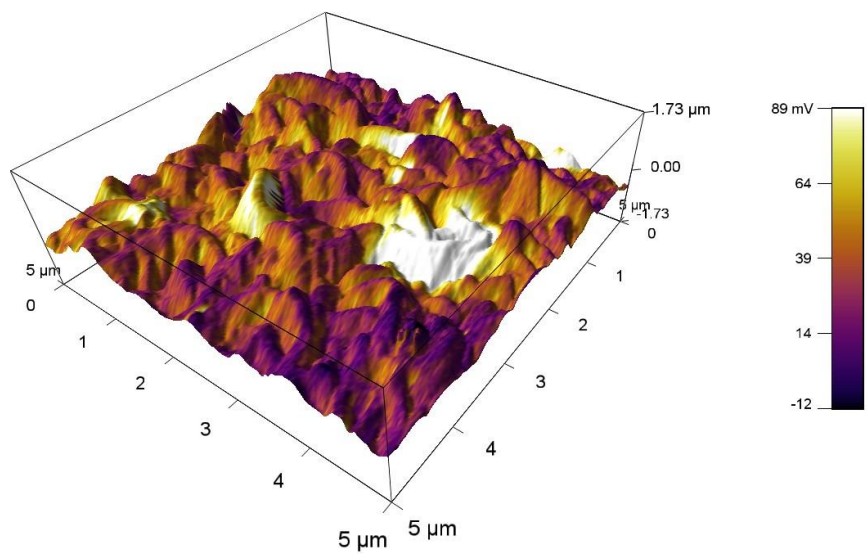

Figure 6. A map of surface capacitance collected from the SCM. The highest scan rate in some regions of the electrode obtained a value of approximately 89mV s$^{-1}$.

The SCM characterization process yielded scan rate values (Figure 6) approximately 35.6 times that of an electrochemical capacitor electrode put under similar tests while lacking the hybrid component of the comparably synthesized device (Lu et al 2012).

Overall, the deposition process of the $TiO_2$ onto the MWCNT/$SiO_2$ proved to be successful and promising as a technique in the design and synthesis of a functional ultracapacitor device. The SCM characterization technique confirmed positive preliminary data about the capacitance and energy density, although more tests are needed to confirm the data.

Discussion

Overall, the synthesis and characterization of an electrochemical capacitor device electrode proved to be successful. The data obtained through the SCM tests proved to be extremely exciting as an improvement over comparable electrochemical capacitance device. More tests are currently underway to expand on the collection of data. In order to expand upon the data of this research, future directions will include the usage of voltammetry methods.  The discharge and capacitance of the electrode material will be tested using cyclic voltammetry and a galvanostatic discharge method. Both will serve to test and compare the electrode material to published values and serve as a good indicator of long-term performance and applicability. While it is difficult at the current time to predict outcomes of this series of tests, it is hoped that the data will provide performance that is significantly better and more applicable than current electrochemical capacitors. Theoretically, this will be the case due to the novel application and compilation of materials that compensate for much of the shortcomings of the current electrochemical capacitors. The application of such a device in the real world is an immensely exciting prospect for the future of energy storage.

Energy from renewable sources has the potential to significantly reduce our dependence on fossil fuels, but when energy is collected from renewable energy sources, it is sent to capacitors, where much of the energy is lost due to inefficiency. Electrochemical capacitors (or more commonly known as supercapacitors) were produced primarily for this purpose, although much of the hype surrounding electrochemical capacitors is still only theoretically possible because of certain limitations such as, the limited surface area of the electrode, the density of the electrode surface, and the inhibition of electro-active components. This research is an important step to hybridize various ideas used in current electrochemical theory to create an energy storage device that is both efficient and inexpensive.

The implementation of such a device is an exciting prospect. With increased efficiency, it may be possible for only a handful of solar panels to power an entire city. An even more exciting prospect is the ability for that kind of power to be accessible everywhere in the world. This research represents an important step forward in developing energy storage devices to meet the energy needs of tomorrow.

Acknowledgements

I express gratitude for the Nanofabrication Facility at Stanford University for allowing access to the deposition equipment and materials and to Dr. Jason Li of Asylum Research for the usage of the Scanning Capacitance Microscopy equipment. I thank the Southern California Academy of Sciences Research Training Program for ongoing guidance through the research process.